**Title:** Evaluating Generative AI as an Educational Tool for Radiology Resident Report Drafting

**Authors**: Antonio Verdone[a,*], Aidan Cardall[b,*], Fardeen Siddiqui[b], Motaz Nashawaty[a], Danielle Rigau[a], Youngjoon Kwon[a], Mira Yousef[a], Shalin Patel[a], Alex Kieturakis[a], Eric Kim[a], Laura Heacock[a], Beatriu Reig[a], Yiqiu Shen[a,c]

**Author Affiliations:**

[a]New York University Grossman School of Medicine, Department of Radiology

[b]New York University Grossman School of Medicine

[c]New York University, Center for Data Science

[*]Equal Contribution

**Corresponding author**: Yiqiu Shen, 227 E 30th St, New York, NY 10016; Yiqiu.Shen@nyulangone.org.

**Keywords**: Generative AI, Large Language Model, Radiology Education, Breast Imaging

**Data Statement**: The author(s) declare(s) that they had full access to all of the data in this study and the author(s) take(s) complete responsibility for the integrity of the data and the accuracy of the data analysis.

**Funding**: This work was supported by the Medical Education Research Grant through NYU Grossman School of Medicine's Program on Medical Education Innovations and Research (PrMEIR) grant program.

**Abstract**

**Objective**: Radiology residents require timely, personalized feedback to develop accurate image analysis and reporting skills. Increasing clinical workload often limits attendings' ability to provide guidance. This study evaluates a HIPAA-compliant GPT-4o system that delivers automated feedback on breast imaging reports drafted by residents in real clinical settings.

**Methods**: We analyzed 5,000 resident–attending report pairs from routine practice at a multi-site U.S. health system. GPT-4o was prompted with clinical instructions to identify common errors and provide feedback. A reader study using 100 report pairs was conducted. Four attending radiologists and four residents independently reviewed each pair, determined whether predefined error types were present, and rated GPT-4o's feedback as helpful or not. Agreement between GPT and readers was assessed using percent match. Inter-reader reliability was measured with Krippendorff's alpha. Educational value was measured as the proportion of cases rated helpful.

**Results**: Three common error types were identified: (1) omission or addition of key findings, (2) incorrect use or omission of technical descriptors, and (3) final assessment inconsistent with findings. GPT-4o showed strong agreement with attending consensus: 90.5%, 78.3%, and 90.4% across error types. Inter-reader reliability showed moderate variability ($\alpha$ = 0.767, 0.595, 0.567), and replacing a human reader with GPT-4o did not significantly affect agreement ($\Delta$ = –0.004 to 0.002). GPT's feedback was rated helpful in most cases: 89.8%, 83.0%, and 92.0%.

**Discussion**: ChatGPT-4o can reliably identify key educational errors. It may serve as a scalable tool to support radiology education.

# Introduction

Developing strong reporting skills is a cornerstone of radiology residency training. Residents are expected not only to interpret imaging accurately but also to communicate findings clearly and consistently through structured reports. Traditionally, this process is supported by iterative feedback: residents draft preliminary reports, and attending radiologists refine them with corrections and teaching points. However, this education paradigm is often disrupted in practice, as increasing clinical workloads have reduced opportunities for timely, iterative feedback, limiting the individualized guidance residents receive [1,2]. This reduced direct engagement can slow the development of essential reporting skills and undermine residents' confidence and consistency, especially for complex imaging cases [3,4].

Large language models (LLMs) offer a promising avenue to bridge this educational gap by providing automated and constructive support in report drafting. Prior studies have demonstrated the potential of LLMs to analyze radiology reports and to support broader medical education initiatives [5,6]. Yet, few efforts have focused on providing targeted, clinically grounded feedback on the actual reports residents generate during routine practice. Existing tools rarely incorporate attending-authored reports as reference standards or operate within authentic clinical workflows. They often lack the ability to deliver precise, actionable guidance on how to improve report quality.

To address this unmet need, we introduce a practical framework that aims to integrate an advanced LLM (GPT-4o [7]) into residency training by comparing resident-drafted reports with the corresponding attending's final version. Our objectives were: (1) to identify common, clinically relevant discrepancies between resident and attending reports from real clinical practice, (2) to assess GPT-4o's reliability in detecting these discrepancies and providing targeted feedback, and (3) to evaluate how radiologists and residents perceive the educational value of its feedback. We hypothesized that GPT-4o could detect key discrepancies with accuracy comparable to expert readers and generate actionable feedback.

## Methods

An overview of the system pipeline and evaluation design is shown in Figure 1. This retrospective study was conducted at a single academic medical center using de-identified radiology reports. This study is structured in accordance with the CLAIM guidelines [8]. All analyses were performed on a HIPAA-compliant, institutionally governed private instance of GPT-4o hosted on Microsoft Azure.

Data Collection

Resident draft and attending final radiology report pairs were collected from the breast imaging section of an anonymized academic institution beginning in June 2023, under institutional data governance protocols. A total of 35,755 pairs were assembled. Eligible studies included mammograms, with or without same-day breast ultrasound, drawn from both screening and diagnostic settings. Breast imaging was selected because of its high clinical workload and reporting complexity. Patients were ≥18 years old, including both women and men. All identifiers were removed prior to analysis.

From this corpus, three datasets were created (Figure 2). First, a random sample of 5,000 pairs was used to identify frequent and clinically meaningful discrepancies between resident and attending reports, forming the Common Error Analysis Set.  Second, we constructed the Reader Study Set to evaluate GPT-4o's performance. From the corpus (excluding cases already used in the Common Error Analysis Set), we randomly sampled 100 report pairs. Pairs were excluded if the resident draft was incomplete (missing findings, impression, or final diagnosis), if the draft and final reports were too similar (Levenshtein distance [9] ≤ 50), or if the study did not include both mammography and same-day breast ultrasound.

The final Reader Study Set comprised 100 cases (99 women and 1 man), with a mean age of 58.9 years (SD = 11.2). The BI-RADS distribution for this set is summarized in Supplementary Table S1. Third, a subset of 15 pairs was selected using the same filtering criteria to represent a balanced mix of the error types, identified through the procedure described in the following subsection. These examples were used to design GPT-4o prompt templates with illustrative inputs and expected outputs, forming the Prompt Sample Set.

## Identifying Common Errors

To identify the most frequent and educationally relevant errors made by residents, we analyzed 5,000 resident–attending report pairs from the Error Analysis Set using GPT-4o in a two-step process. First, we designed a prompt that included both the resident's draft and the attending's finalized report. GPT-4o was asked to summarize the differences in findings and diagnosis between the residents' drafts and attendings' finalized reports for each pair. This process was automated via API queries across all 5,000 pairs. Second, the resulting 5,000 summaries were compiled into a single input, and GPT-4o was prompted to extract the most common error types along with their frequencies. The complete prompts are provided in Supplementary Section S1. The three most frequent error types were selected for further analysis.

## LLM Feedback System

We developed two prompts to guide GPT-4o in identifying the three most common error types in resident–attending report pairs: Inconsistent Findings, Inconsistent Descriptions, and Inconsistent Diagnoses. Each prompt instructed GPT-4o to make a binary (Yes/No) judgment on whether each of these three errors was present and to provide explanatory feedback in all cases, including why the resident's draft was incorrect when errors were identified.

Two prompts were developed. The first prompt was used for detecting Inconsistent Findings and Inconsistent Descriptions. It included both the resident's draft and the attending's finalized report, along with structured comparison guidance, a complete BI-RADS lexicon, BI-RADS score definitions and criteria, and 15 illustrative output examples. The second prompt focused on identifying Inconsistent Diagnoses. It analyzed only the resident's draft to assess whether the assigned BI-RADS category was supported by the descriptive content in the findings and impressions. Full prompt texts are provided in Supplementary Section S2. Illustrative examples of GPT's comments in response to these prompts are presented in Supplementary Section S3.

## Reader Study

To evaluate the reliability and educational value of GPT-4o's feedback, we conducted a reader study with eight participants: four attending radiologists, all of whom are board-certified, fellowship-trained in breast imaging with 2, 4, 5, and 15 years of post-fellowship experience, and four radiology residents (three first-year and one second-year). Each reader independently reviewed the same set of 100 report pairs from the Reader Study Set.

For each report pair, readers indicated with a binary response (Yes/No) whether each of the three predefined errors was present. GPT-4o performed the same evaluations in parallel using the prompts described above. After completing their independent assessments (blinded to GPT's output), readers were shown GPT-4o's responses and accompanying explanations. They were then asked to judge the usefulness of GPT-4o's feedback by selecting either "Yes, helpful" or "No, not helpful," to each of the three error types, with an option to leave additional comments. The entire study was conducted using a custom-built interface in Label Studio, illustrated in Figure S1.

## Statistical Analysis

We assessed agreement between GPT-4o and each reader, as well as with the attending majority consensus, using Cohen's kappa (κ), exact agreement rate (i.e., the proportion of cases where both GPT and the reader gave the same Yes/No judgment), precision, recall, and F1-score. Cohen's kappa values were interpreted using established thresholds: 0.41 – 0.60 = moderate agreement, 0.61 – 0.80 = substantial agreement [10]. Ninety-five percent confidence intervals (CIs) were estimated using 10,000 bootstrap iterations.

To evaluate inter-rater reliability among the eight readers, we used Krippendorff's alpha (α) [11], which is a robust measure of agreement that accounts for chance agreement, accommodates binary and missing data, and does not assume independent or interchangeable raters. To quantify the impact of including GPT-4o on inter-reader agreement, we replaced each human reader, one at a time, with GPT-4o and recalculated α for each configuration. This resulted in eight new α values, each reflecting the agreement level when GPT-4o replaced a specific reader. We then compared each of these to the original α among all eight readers to compute the difference in agreement. The average of these differences, denoted by Δ, defines the effect size: a positive Δ indicates that GPT-4o generally improved inter-reader agreement, while a negative Δ indicates a decrease. We estimated 95% CIs for Δ via bootstrap and conducted a two-sided permutation test (10,000 iterations) to determine whether the observed change is statistically significant.

Finally, we evaluated perceived usefulness by asking readers to review GPT-4o's feedback and explanations for each case and indicate whether the feedback was helpful. We reported the proportion of "Yes, helpful" responses and stratified results by reader experience (attending vs. resident).

### Qualitative Analysis of Reader Comments

Free-text comments were thematically analyzed to characterize perceived limitations of GPT's feedback. All comments were compiled in a spreadsheet and manually reviewed. Comments were grouped into thematic categories based on recurring content, and each was labeled with a descriptive theme (e.g., "Incorrect Answer," "Stylistic differences"). Frequencies were calculated to summarize common themes.

## Results

### Common Errors

The most prevalent errors included unclear or ambiguous descriptions (35%), inconsistent use of medical terminology (32%), and omission of key imaging findings present in the attending's report (28%). Less frequent issues included missing or incorrect BI-RADS assessments (12%), inappropriate or absent follow-up recommendations (8%), failure to incorporate relevant clinical history (5%), and a general lack of structured reporting (5%). From these patterns, we defined three clinically meaningful and automatable error categories for downstream analysis: (1) Inconsistent Findings – omission or addition of significant findings relative to the attending report; (2) Inconsistent Descriptions – misuse or omission of BI-RADS lexicon terms; and (3) Inconsistent diagnoses – BI-RADS score not supported by the resident's own description.

### Reliability Assessment

GPT-4o showed moderate to substantial agreement with the attending radiologist's consensus across all three error types. For Inconsistent Findings, it achieved 90.5% agreement (95% CI: 84.2%–95.8%) and a Cohen's κ of 0.790 (95% CI: 0.647–0.908), indicating substantial agreement. For Inconsistent Descriptions, agreement was moderate (κ = 0.550, 95% CI: 0.368–0.716) with 78.3% agreement (95% CI: 69.6%–87.0%). For Inconsistent Diagnoses, agreement was substantial (κ = 0.615, 95% CI: 0.356–0.825) with 90.4% agreement (95% CI: 84.0%–95.7%). Table 1 provides a detailed summary of the statistical results. Examples of cases where the reader and GPT diverged are presented in Figure S2.

## Impact of GPT on Inter-Reader Agreement

Inter-reader variability was evident across the three error types (Table S2). For Inconsistent Findings, readers showed substantial agreement (α = 0.767, 95% CI: 0.679–0.842). However, agreement declined for Inconsistent Descriptions (α = 0.595, 95% CI: 0.506–0.680) and Inconsistent Diagnoses (α = 0.567, 95% CI: 0.395–0.695). In addition, for all three error types, attending radiologists demonstrated higher inter-rater agreement than residents.

As shown in Figure 3, adding GPT-4o to the reader panel resulted in minimal and statistically insignificant changes in agreement for identifying all three error types (Inconsistent Findings: Δ = -0.004, p = 0.875; Inconsistent Description: Δ = -0.013, p = 0.626; Inconsistent Diagnoses: Δ = 0.002, p = 0.751). This pattern remained consistent in subgroup analyses of attendings and residents, indicating that GPT-4o did not significantly affect inter-reader reliability. Detailed agreement metrics are provided in Supplementary Table S2.

## Perceived Helpfulness of GPT-4o Feedback

GPT-4o's feedback was rated helpful in 86.83% (95% CI: 84.21%–89.21%) of all evaluations across readers and error types. Among attendings, it was considered helpful in 87.75% (95% CI: 84.50%–90.25%) of cases for Inconsistent Findings, 81.50% (95% CI: 73.00%–89.00%) for Inconsistent Descriptions, and 87.00% (95% CI: 83.50%–90.00%) for Inconsistent Diagnoses. Residents rated GPT's feedback more favorably: 89.75% (95% CI: 87.00%–92.50%), 83.00% (95% CI: 78.25%–89.75%), and 92.00% (95% CI: 88.00%–96.00%), respectively, with the greatest difference in Inconsistent Diagnoses. This suggests that GPT's structured explanations are especially beneficial for less experienced readers, particularly in evaluating BI-RADS justification.

## Qualitative Analysis of Reader Comments on GPT-4o Responses

Of the 311 total comments, 266 were made when GPT-4o was rated unhelpful, suggesting readers were more inclined to comment on unsatisfactory feedback. Among these, 228 comments occurred when GPT-4o disagreed with the reader, typically reiterating its errors, while only 38 were made when GPT-4o agreed. Among all three error types, Inconsistent Description prompted the most comments. Supplementary Table S3 summarizes comments by the reader group and error types.

Qualitative analysis included only comments that directly addressed GPT-4o; unclear or unrelated remarks were excluded. As shown in Figure 4, four major themes emerged:

1. **Error Type Confusion (21.7%):** GPT-4o misclassified the error type, often conflating Inconsistent Findings with Inconsistent Description, even when its judgment aligned with the reader.

2. **Incorrect Answer (63.8%):** GPT-4o disagreed with the reader, and readers explained why it was wrong.

3. **Stylistic Differences (10.1%):** GPT-4o is overly strict about minor stylistic differences or interchangeable terms.

4. **Clinical Irrelevance (4.3%):** GPT-4o identified issues that were not clinically meaningful.

Notably, two comments acknowledged that GPT-4o's answer was correct but the rationale was flawed, while another two noted that GPT-4o helped readers recognize previously missed details. Representative examples are provided in Supplementary Section S4.

**Discussion**

This study evaluated the use of GPT-4o as an educational tool to improve radiology residents' report drafting skills. Our goals were to: (1) identify common reporting errors made by residents, (2) develop an LLM-based system to automatically detect these errors and provide targeted feedback, and (3) assess the system's reliability and perceived educational value. We focused on three error types—Inconsistent Findings, Inconsistent Descriptors, and Inconsistent Diagnoses—defined through LLM analysis of paired reports from routine clinical practice, each consisting of a resident's draft and the attending radiologist's finalized report for the same case, collected from routine clinical practice. In a reader study, the system achieved reliability comparable to radiologists, and its feedback was rated as helpful by both residents and attendings.

The ACGME recognizes reporting as a core patient care competency for radiology residents [12], with the goal of producing efficient, structured reports that use appropriate lexicons to meet both provider and subspecialty needs. Traditionally, reporting skills are taught by attendings during daily workflow and in didactic sessions. However, since 2006, rising clinical workloads have reduced the time available for attendings to provide this instruction, competing with the demands of teaching image interpretation and other key competencies [13]. An automated feedback tool could help alleviate this burden, enabling attendings to focus on other essential training needs.

AI offers a promising way to meet this need. In medical education, AI has been applied in areas such as intelligent tutoring, automated assessment, and personalized learning pathways [14–18]. In radiology, AI has been used to tailor educational content, provide automated image-based feedback [19–23], and deliver case-specific recommendations during reading sessions [24]. LLMs have also been used to summarize radiology reports, translate medical jargon, and extract key findings [25,26]. However, prior work rarely provides targeted, clinically grounded feedback on residents' real-world report drafts, and most lack actionable guidance for improving report quality. Our study addresses this gap by enabling GPT-4o to analyze paired resident and attending reports from actual clinical practice—a data source not previously explored—to deliver personalized, case-specific feedback focused on common real-world error types.

Unlike commercially available reporting systems [27] that simply highlight textual differences and compare revisions to the resident's preliminary report within the final report, our AI tool goes beyond simple text comparison by identifying and categorizing clinically relevant errors and explaining each error. Additionally, the LLM solution filters out insignificant edits, such as purely stylistic changes like reordering phrases, making the review process more efficient and focused on meaningful learning for residents.

This AI tool could be deployed in three main ways. First, it can provide daily or weekly automated feedback, analyzing all resident–attending report pairs from routine clinical work and delivering personalized, case-specific guidance that residents can review outside the time-pressured reading room. Second, it can support targeted educational sessions by aggregating error patterns across residents, enabling faculty to highlight common issues and address them in small-group teaching or case conferences using anonymized real cases. Third, it can facilitate competency tracking and progress assessment by storing longitudinal feedback for each resident, allowing program directors to monitor improvement in reporting skills, identify persistent weaknesses, and tailor remediation plans accordingly.

Despite these promising results, our study has several limitations. First, the AI system showed the greatest disagreement with readers in detecting Inconsistent Descriptor errors, which was also the area of highest disagreement among readers themselves. Error analysis suggested that most disagreements stemmed from deciding whether changes in technical descriptors were clinically significant or merely stylistic. As a large institution with 18 teaching attendings in breast imaging, reporting styles vary: some specify the size and location of all cysts, while others use general statements such as "benign cysts in both breasts"; some include benign findings (e.g., biopsy clips, benign calcifications), while others report only new or significant findings. When such differences do not alter the BI-RADS diagnosis or recommendation, their significance is subjective. Given the moderate inter-reader reliability in this area, improving GPT agreement with attendings may be challenging. A potential solution is to train GPT with examples illustrating stylistic variability so it can flag differences as possibly subjective. Second, in this study, attending reports were used as the gold standard without re-reviewing the imaging, yet these reports may also contain variability. For example, we observed occasional use of non-BI-RADS terms (e.g., "nodule" instead of "mass") or outdated BI-RADS terminology ("cluster" instead of "group"). This suggests an additional application of our system—promoting consistent BI-RADS lexicon use in attending reports. Third, our evaluation was limited to structured breast imaging reports from a single center. The generalizability of the system to less structured notes, other pathologies, or different imaging modalities remains unknown.

Future work could focus on multi-center validation, expansion to other body parts and modalities, and randomized controlled trials to assess the tool's longitudinal educational impact. Ultimately, this LLM-powered system offers a scalable, clinically grounded, and personalized approach to enhancing radiology resident training, addressing a critical need for timely, high-quality feedback in an era of increasing clinical demands.

# Figures

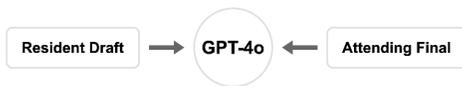

**Figure 1:** Schematic overview of the system and evaluation. (A) Data collection and error type identification. The report pairs shown are oversimplified examples for demonstration purposes. (B) GPT-4o feedback system. (C) Reader study.

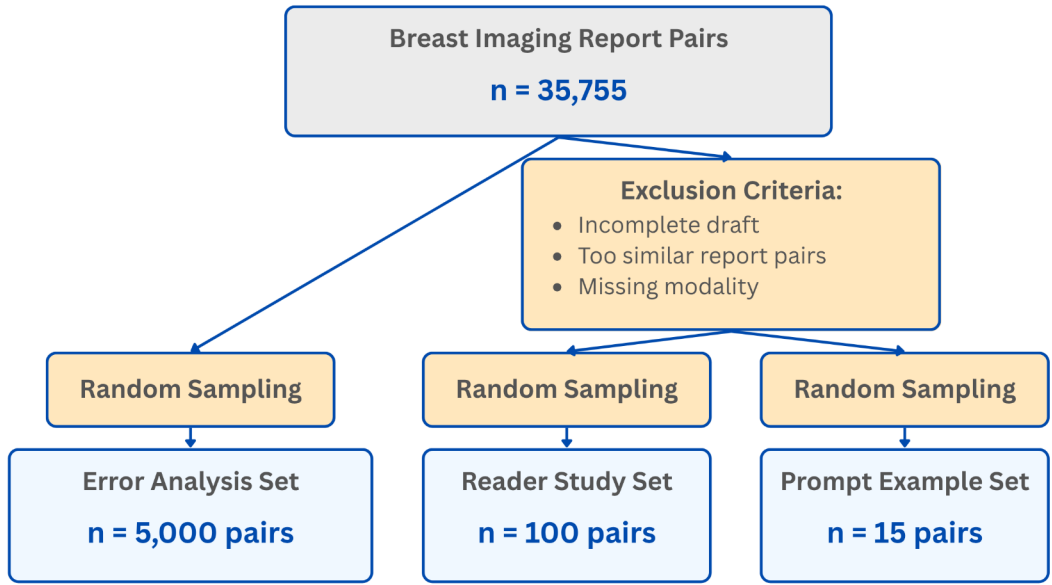

**Figure 2:** A total of 35,755 resident–attending report pairs were collected from the breast imaging section at an anonymized academic medical center. From this set, 5,000 pairs were randomly sampled for GPT-4o to analyze common error patterns. Additional exclusion criteria were applied to the remaining pairs: incomplete resident drafts, pairs with minimal differences (edit distance < 50), and cases lacking either mammography or ultrasound reports were removed. From the filtered dataset, 100 pairs were selected for the reader study and 15 additional pairs were used as instructional examples in GPT's prompt.

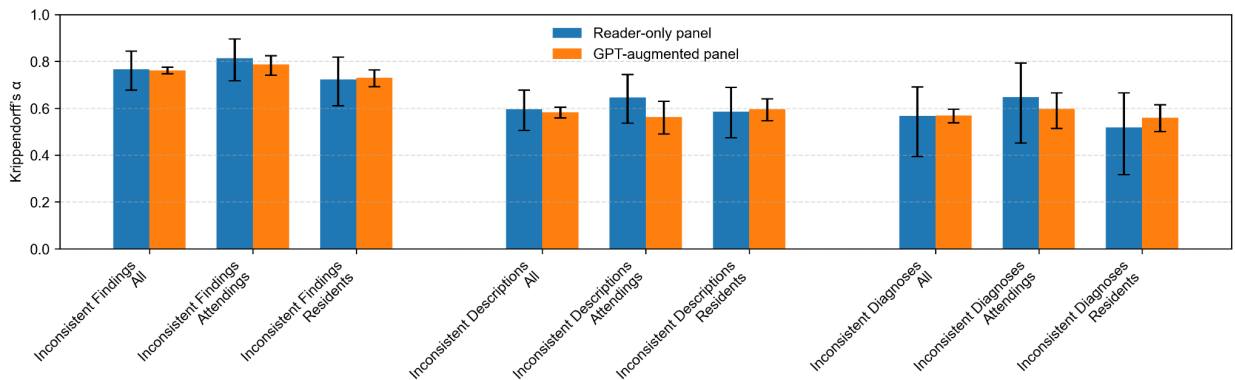

**Figure 3:** Krippendorff's alpha for each error type (Inconsistent Finding, Inconsistent Description, Inconsistent Diagnoses), comparing inter-reader agreement among the original human-only panel and GPT-augmented panels. For each error type, GPT-4o individually replaced one of the eight human readers, and Krippendorff's alpha was recalculated. Values shown for the GPT-augmented group represent the average α across all substitutions. Error bars indicate 95% confidence intervals obtained via bootstrap resampling. The narrower intervals in the GPT-augmented group reflect reduced variability due to averaging across multiple substitutions.

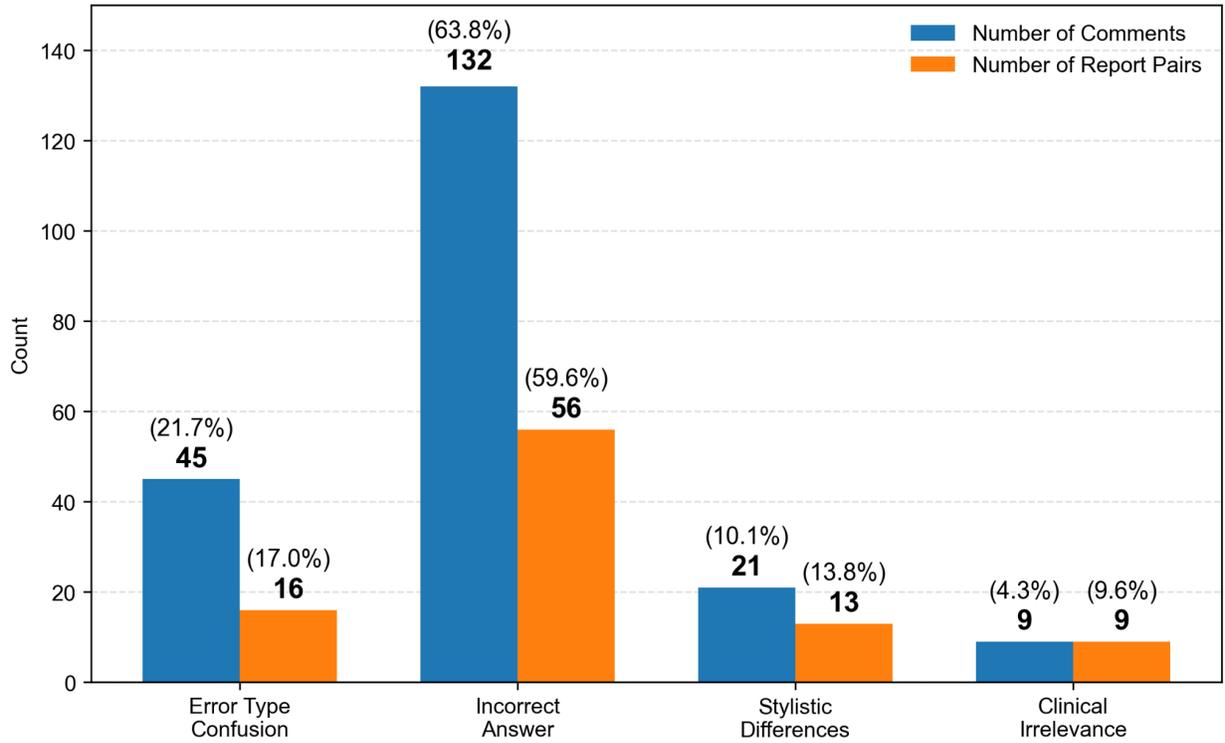

**Figure 4:** Distribution of themes in reader comments on GPT-4o's feedback. The chart displays the number and percentage of comments, as well as the number of unique report pairs, associated with each identified theme. Note that multiple comments could be made for the same report pair.

# Tables

**Table 1**: GPT-4o's Agreement with Attending Radiologist Consensus. This table summarizes the agreement between GPT-4o and the consensus of four attending radiologists on report pairs from clinical practice (ties excluded). Columns show: (1) number of consensus report pairs (N), (2) percentage of pairs labeled as containing an error by the attending consensus ("Reader Yes"), (3) percentage labeled by GPT as containing an error ("GPT Yes"), (4) overall agreement rate between GPT and the consensus, and (5) Cohen's kappa measuring agreement beyond chance. Higher values indicate stronger alignment.

|  | N | Reader "Yes" | GPT "Yes" | Percentage of Agreement | Cohen's κ |
|---|---|---|---|---|---|
| **Inconsistent Findings** | 95 | 33.50% | 38% | 90.5% (84.2%–95.8%) | 0.790 (0.647–0.908) |
| **Inconsistent Descriptions** | 92 | 38.75% | 44% | 78.3% (69.6%–87%) | 0.550 (0.368–0.716) |
| **Inconsistent Diagnoses** | 94 | 15% | 20% | 90.4% (84%–95.7%) | 0.615 (0.356–0.825) |

# Supplementary Materials

S1: Prompts to Identify Common Error Types

Prompt used to extract differences between report a pair of resident's draft and radiologist's finalized report:

> You are an experienced breast radiologist. I will provide two radiology reports (one from the attending and one from the resident) which could contain mammograms, ultrasound, or both. Your task is to summarize the differences in findings and diagnosis between the two radiology reports. Be concise and use bullet points.
>
> Attending's finalized report: {to_be_inserted}
>
> Resident's draft report: {to_be_inserted}

Prompt used to summarize the differences into common mistakes made by residents:

> Please analyze the provided list of discrepancies in findings and diagnoses between two radiology reports, which may include mammography, ultrasound, or both. Summarize the most common types of discrepancies you observe and report their frequencies.
>
> 1, difference in findings
>
> 2, difference in findings
>
> …

# S2: Prompts to Detect Error Types in Report Pairs

Prompts to detect the three defined error types in resident-attending report pairs. Each prompt contains 15 example reports, however for brevity we are showing only 1.

Prompt for *Inconsistent findings* and *Inconsistent descriptions*:

> Please follow the instructions provided in Part 1: Task Statement, using the reference materials, including "Part 2: BI-RADS lexicon terms for describing breast masses", "Part 3: qualifying criteria for the proposed BI-RADS score", and "Part 4: illustrative examples with chain-of-thought explanations". The input to be processed is specified in "Part 5: Request" at the end of this prompt.
>
> PART 1: Task Statement I will provide you with a pair of breast radiology reports. One report is an attending's report and the other is a resident's report. A radiology report may include a breast ultrasound, a mammogram, or a combined ultrasound and mammogram. The attending's report represents the standard for a well-structured radiology report. I want you to identify discrepancies or inconsistencies between the attending's report and the resident's report. You need to look for the 2 following discrepancies or inconsistencies:
>
> 1. Inconsistent Findings: Compare the resident's report with the attending's report and look for findings that are either present in the attending's report but missing from the resident's report, or present in the resident's report but not mentioned in the attending's report. Note: This category focuses on the presence or absence of findings, not the way they are described. If this happens, return True. Otherwise, return False. Also, explain your choice.

2. Inconsistent Descriptions: Compare the resident's report with the attending's report for all findings included in the resident's report. Look for instances where the resident's report omits BI-RADS lexicon descriptors that are specified in the attending's report, or where the resident's report uses informal or inaccurate descriptors that are corrected or replaced by appropriate BI-RADS terms in the attending's report. Note: This category focuses on the accuracy and completeness of the descriptions of findings that are present in both reports. If this happens, return True. Otherwise, return False. Also, explain your choice.

Here is the process for answering the aforementioned questions:

1. Extract Findings and BI-RADS Score: Extract all findings and the BI-RADS score from both the resident's report and attending's report.

2. Compare Findings: Align and compare findings between the resident's report and the attending's report. Identify discrepancies to answer the "Inconsistent Findings" question.

3. Compare Descriptions: For each finding in the resident's report, compare its description with the corresponding finding in the attending's report. Refer to Part 2 for the BI-RADS lexicon to identify any missing or misused descriptors and answer the "Inconsistent Descriptions" question.

4. Report Discrepancies: For all 2 questions, if any discrepancy is found, mark "True" and provide an explanation. Otherwise, mark "False."

5. Generate Output: Present findings and explanations in a csv-like format, separating each item with a vertical bar "|".

PART 2: Here is a dictionary of BI-RADS lexicon terms to describe breast masses:

"Tissue composition: fat homogeneous background echotexture, fibroglandular homogeneous background echotexture, heterogeneous background echotexture.

Mass shape: oval, round, irregular.

Orientation: parallel, not parallel.

Mass margin: circumscribed, angular, microlobulated, indistinct, spiculated.

Echo pattern: anechoic, hyperechoic, hypoechoic, isoechoic, heterogeneous, complex cystic and solid.

Posterior features: No posterior features, enhancement, shadowing, combined pattern.

Calcifications: calcifications in a mass, calcifications outside of a mass, intraductal calcifications.

Calcification morphology: fine linear or fine-linear branching, fine pleomorphic, amorphous, coarse heterogeneous, punctate, skin, vascular, coarse or "popcorn-like", large rod-like, round, rim (historically "eggshell"), dystrophic, milk of calcium, suture.

Calcification distribution: segmental, linear, grouped, regional, diffuse.

Associated features: Architectural distortion, duct changes, skin thickening, skin retractions, edema.

Vascularity: Absent, internal vascularity, vessels in rim.

Special cases: simple cyst, clustered microcysts, complicated cyst, mass in or on skin, foreign body including implants, intramammary lymph nodes, axillary lymph nodes, vascular abnormalities, arteriovenous malformations/pseudoaneurysms, Mondor disease, postsurgical fluid collection, fat necrosis."

PART 3: Here are the qualifying criteria for the proposed BI-RADS score:

"BI-RADS 0 Need Additional Imaging Evaluation: Utilized when further imaging evaluation (additional views or ultrasound) or retrieval of prior examinations is required.

BI-RADS 1 Negative: There is nothing to comment on. The breasts are symmetric and no masses, architectural distortion or suspicious calcifications are present. Use BI-RADS 1 if there are no abnormal imaging findings in a patient with a palpable abnormality, but add a sentence recommending surgical consultation or tissue diagnosis if clinically indicated.

BI-RADS 2 Benign Finding: Like BI-RADS 1, this is a normal assessment, but here, the radiologist chooses to describe a benign finding like:

* Follow up after breast conservation surgery * Involuting, calcified fibroadenomas

* Multiple large, rod-like calcifications

* Rim or oil cyst calcifications

* Layering calcifications or milk of calcium * Intramammary lymph nodes

* Vascular calcifications

* Implants

* Architectural distortion clearly related to prior surgery.

* Fat-containing lesions such as oil cysts, lipomas, galactoceles and mixed-density hamartomas. * Abscess or hematoma

They all have characteristically benign appearances, and may be labeled with confidence.

Use in screening or in diagnostic imaging when a benign finding is present. Use in the presence of bilateral lymphadenopathy, probably reactive or infectious in origin

BI-RADS 3 Probably Benign Finding: A finding placed in this category should have less than a 2% risk of malignancy. Lesions appropriately placed in this category include:

* Non-calcified circumscribed mass on a baseline mammogram (unless it can be shown to be a cyst, an intramammary lymph node, or another benign finding),

* Focal asymmetry which becomes less dense or partially effaces on spot compression view

* Solitary group of punctate or round calcifications

* Probable abscess or hematoma

Use in findings on mammography like:

* Noncalcified circumscribed solid mass

* Focal asymmetry

* Solitary group of punctuate or round calcifications

Use in findings on US with robust evidence to suggest * Typical fibroadenoma

* Isolated complicated cyst

* Clustered microcysts

BI-RADS 4 or 5 Suspicious Abnormality or Highly Suggestive of Malignancy: These categories are reserved for findings that are sufficiently suspicious to justify a recommendation for biopsy. BI-RADS 4 has a wide range of probability of malignancy (2 - 95%). BI-RADS 5 has a >95% likelihood of malignancy. Use BI-RADS 4 or 5 in findings such as:

* Mass without classic features of fibroadenoma or complicated cyst

* Complex cystic and solid mass

* Probable chronic granulomatous mastitis

* Group of calcifications without benign or probably benign features * Mass with irregular shape.

* Spiculated margin.

* High density.

* Ultrasound also shows irregular shape with indistinct margin.

BI-RADS 6: Known biopsy proven malignancy. Use after incomplete excision Use after monitoring response to neoadjuvant chemotherapy."

Part 4 : Illustrative Examples.

Example 1 INPUT :

Attending's report:  MAMMO TOMOSYNTHESIS SCREENING BILATERAL, US BREAST COMPLETE BILATERAL    Clinical Breast Exam: Patient does report clinical breast exam in the last year.    Clinical Indication: Routine screening. Family history of maternal aunt with breast cancer.    Compared to: \*\*/\*\*/\*\*\*\* screening mammogram and ultrasound, \*\*/\*\*/\*\*\*\* screening ultrasound, \*\*/\*\*/\*\*\*\* screening mammogram, \*\*/\*\*/\*\*\*\* screening mammogram and ultrasound    MAMMOGRAM: Tomosynthesis 3D and 2D imaging of the breast(s) were performed.  Current study was also evaluated with a computer aided detection (CAD) system.    Breast density: Heterogeneously dense, which may obscure small masses.    There is questioned distortion in the right breast 9:00 middle depth, best seen on MLO slice 30/73 and CC slice 32/73. This may represent crossing vessels but is indeterminate.  No suspicious masses, calcifications, or other findings are seen in the left breast.    US BREAST COMPLETE BILATERAL    Ultrasound evaluation was performed including examination of all four quadrants of the breast(s) and the retroareolar region(s).    No suspicious abnormalities were seen sonographically.    There are scattered bilateral benign-appearing cysts.    Mammo BI-RADS 0: INCOMPLETE - NEED ADDITIONAL IMAGING EVALUATION    Ultrasound BI-RADS 2: BENIGN    IMPRESSION:    Indeterminate right breast asymmetry with questioned distortion. A right diagnostic mammogram with additional views and possible targeted ultrasound is recommended.    OVERALL BI-RADS 0: INCOMPLETE    An additional imaging exam of the right breast(s) is recommended.    The patient will be sent a letter to return for additional imaging.

Resident's report: MAMMO TOMOSYNTHESIS SCREENING BILATERAL, US BREAST COMPLETE

BILATERAL   Clinical Breast Exam: Patient does report clinical breast exam in the last year.    Clinical Indication: Routine screening. Family history of maternal aunt with breast cancer.    Compared to: \*\*/\*\*/\*\*\*\* screening mammogram and ultrasound, \*\*/\*\*/\*\*\*\* screening ultrasound, \*\*/\*\*/\*\*\*\* screening mammogram, \*\*/\*\*/\*\*\*\* screening mammogram and ultrasound    MAMMOGRAM: Tomosynthesis 3D and 2D imaging of the breast(s) were performed.  Current study was also evaluated with a computer aided detection (CAD) system.    Breast density: Heterogeneously dense, which may obscure small masses.    No suspicious masses, calcifications, or other findings are seen.     US BREAST COMPLETE BILATERAL    Ultrasound evaluation was performed including examination of all four quadrants of the breast(s) and the retroareolar region(s).    No suspicious abnormalities were seen sonographically. There are scattered bilateral benign cysts.  -Right 5:00 2 cm FN 1.2 cm transverse, 1.1 x 0.5 cm (previously 1.3 cm transverse, 1.4 x 0.6 cm) oval isoechoic mass is decreased in size and benign    Mammo BI-RADS 1: NEGATIVE    Ultrasound BI-RADS 2: BENIGN    IMPRESSION:     No imaging evidence of malignancy on the current exam(s).     OVERALL BI-RADS 2: BENIGN    A 1 year screening mammogram and ultrasound of both breast(s) is recommended.    The patient will be sent a normal letter.

Example 1 OUTPUT :

Inconsistent Finding: True |
Explanation: The attending's report notes a distortion in the right breast at 9:00 (middle depth), visible on MLO slice 30/73 and CC slice 32/73, which is absent in the resident's report. In the ultrasound, the resident provided detailed descriptions of scattered bilateral benign cysts, while the attending gave a more general mention. Despite consistency in ultrasound findings, the overall findings are

> inconsistent due to the missing mammogram detail in the resident's report. |
>
> Inconsistent Description: False |
>
> Explanation: The resident's report includes no mammogram findings but uses correct BI-RADS language for the ultrasound descriptions, indicating no inconsistency in descriptions.
>
> Part 5: Request
>
> Please analyze the following reports:
>
> Attending physician's report: {raw_report_final}
>
> Radiology resident's report: {raw_report_draft}

Prompt for *Inconsistent Diagnoses*:

> You are an expert radiologist. Your task is to identify inconsistencies in the BI-RADS scores of breast reports. A report may include a breast ultrasound, a mammogram, or a combined ultrasound and mammogram. You need to analyze independently the BI-RADS scores from the ultrasound, the BI-RADS from the mammogram and the overall BI-RADS.
>
> Follow these steps to analyze the report:
>
> Step 1 – Extract the findings corresponding to the ultrasound part of the report.
>
> Step 2 – Extract the findings corresponding to the mammogram part of the report.

Step 3 – Extract the BI-RADS score of the ultrasound part of the report.

Step 4 – Extract the BI-RADS score of the mammogram part of the report.

Step 5 – Extract the overall BI-RADS score of the report. If the overall BI-RADS is not explicitly mentioned, the overall BI-RADS is the last BI-RADS in the report.

Step 6 – Analyze the findings and the BI-RADS score of the ultrasound part of the report and determine whether there are inconsistencies or not. Analyze also the follow-up recommendations against the BI-RADS score.

Step 7 – Analyze the findings and the BI-RADS score of the mammogram part of the report and determine whether there are inconsistencies or not. Analyze also the follow-up recommendations against the BI-RADS score.

Step 8 – Analyze whether the overall BI-RADS is consistent with the ultrasound and mammogram BI-RADS. The overall BI-RADS score should be the same as the most severe score between the ultrasound and mammogram. Ensure that the overall BI-RADS score is consistent with the findings and recommendations provided in the report.

Step 9 – Return True (inconsistencies exist) or False (no inconsistencies), along with a brief explanation. Generate the output in a CSV-like format, with the format: Inconsistent BI-RADS: Boolean

Decision | Explanation: explanation. Analyze your explanation so that it is consistent with your boolean decision.

Here are the qualifying criteria for the proposed BI-RADS score:

- BI-RADS 0 Need Additional Imaging Evaluation: Utilized when further imaging evaluation (additional views or ultrasound) or retrieval of prior examinations is required.
- BI-RADS 1 Negative: There is nothing to comment on. The breasts are symmetric and no masses, architectural distortion or suspicious calcifications are present. Use BI-RADS 1 if there are no abnormal imaging findings in a patient with a palpable abnormality, but add a sentence recommending surgical consultation or tissue diagnosis if clinically indicated.
- BI-RADS 2 Benign Finding: Like BI-RADS 1, this is a normal assessment, but here, the radiologist chooses to describe a benign finding like the following ones:

    * Follow up after breast conservation surgery.

    * Involuting, calcified fibroadenomas.

    * Multiple large, rod-like calcifications.

    * Rim or oil cyst calcifications.

    * Layering calcifications or milk of calcium.

    * Intramammary lymph nodes.

    * Vascular calcifications.

    * Implants.

    * Architectural distortion clearly related to prior surgery.

    * Fat-containing lesions such as oil cysts, lipomas, galactoceles and mixed-density hamartomas.

* Abscess or hematoma.

They all have characteristically benign appearances, and may be labeled with confidence.

Use in screening or in diagnostic imaging when a benign finding is present. Use in the presence of bilateral lymphadenopathy, probably reactive or infectious in origin

- BI-RADS 3 Probably Benign Finding: A finding placed in this category should have less than a 2% risk of malignancy. Lesions appropriately placed in this category include:

* Non-calcified circumscribed mass on a baseline mammogram (unless it can be shown to be a cyst, an intramammary lymph node, or another benign finding).

* Focal asymmetry which becomes less dense or partially effaces on spot compression view.

* Solitary group of punctate or round calcifications.

* Probable abscess or hematoma.

Use in findings on mammography like:

* Noncalcified circumscribed solid mass.

* Focal asymmetry.

* Solitary group of punctuate or round calcifications.

Use in findings on ultrasound with robust evidence to suggest:

* Typical fibroadenoma

* Isolated complicated cyst

* Clustered microcysts

A 3-months or 6-months follow-up is associated with a BI-RADS 3.

- BI-RADS 4 or 5 Suspicious Abnormality or Highly Suggestive of Malignancy: These categories are reserved for findings that are sufficiently suspicious to justify a recommendation for biopsy. BI-RADS 4 has a wide range of probability of malignancy (2 - 95%). BI-RADS 5 has a >95% likelihood of

malignancy. Use BI-RADS 4 or 5 in findings such as:

* Mass without classic features of fibroadenoma or complicated cyst.

* Complex cystic and solid mass.

* Probable chronic granulomatous mastitis.

* Group of calcifications without benign or probably benign features.

* Mass with irregular shape.

* Spiculated margin.

* High density.

* Ultrasound also shows irregular shape with indistinct margin.

- BI-RADS 6: Known biopsy proven malignancy. Use after incomplete excision Use after monitoring response to neoadjuvant chemotherapy.

Here are some examples:

Example 1 INPUT

Report: MAMMO TOMOSYNTHESIS SCREENING BILATERAL    Clinical Breast Exam: Patient does report clinical breast exam in the last year.    Clinical Indication: Routine screening. No family history of breast cancer.   Comparison: None    MAMMOGRAM:    Tomosynthesis 3D and 2D imaging of the breast(s) were performed.  Current study was also evaluated with a computer aided detection (CAD) system.   Breast density: Heterogeneously dense, which may obscure small masses.    No suspicious masses, calcifications, or other findings are seen in the right breast. There is a mass in the inner middle depth left breast.    Mammo BI-RADS 0: INCOMPLETE - NEED ADDITIONAL IMAGING EVALUATION     US

> BREAST COMPLETE BILATERAL    Ultrasound evaluation was performed including examination of all four quadrants of the breast(s) and the retroareolar region(s).    No suspicious abnormalities were seen sonographically. Benign 5 mm cyst in the right breast.    Ultrasound BI-RADS 2: BENIGN OVERALL IMPRESSION: OVERALL BI-RADS 0: INCOMPLETE    Mass in the inner left breast, recommend spot magnification view.    An additional imaging exam of the left breast(s) is recommended.    The patient will be sent a letter to return for additional imaging.
>
> Example 1 OUTPUT
>
> Inconsistent BI-RADS: False |
>
> Explanation: The report BI-RADS aligns with the reported finding of mass in the left breast. Thus, there is no inconsistency in the BI-RADS assignment.
>
> Analyze the following report: {raw_report_draft}

## S3: GPT's Feedback Examples

Illustrative comments generated by GPT-4o in cases where discrepancies were detected in response to the three prompts.

**Feedback for *Inconsistent findings:***

- The attending's report mentions "left retroareolar cysts" in the ultrasound findings, which are absent in the resident's report. This discrepancy indicates inconsistent findings between the two reports.

**Feedback for *Inconsistent descriptions:***

- The resident's report describes the breast density as "heterogeneously dense," while the attending's report describes it as "scattered areas of fibroglandular density." The resident's description does not align with the BI-RADS lexicon term used by the attending.

**Feedback for *Inconsistent diagnoses:***

- The report assigns an overall BI-RADS score of 2 (Benign) despite describing a focal asymmetry in the left breast as probably benign, which typically corresponds to BI-RADS 3. The follow-up recommendation of 12 months for the left focal asymmetry is inconsistent with the BI-RADS 2 score, as BI-RADS 3 findings usually warrant a 6-month follow-up. Therefore, the BI-RADS score does not align with the findings and recommendations described in the report.

## S4: Reader Comments on GPT's Feedback

Example reader comments about GPT-4o's responses, illustrating each major theme identified in the qualitative analysis.

**Question Confusion**

- "Description incorrect, but likely same finding"
- "chatgpt confuses a type 1 and type 2 finding because of the typo in the report; it's actually the same finding"
- "Findings are consistent in each report, only descriptors differ"
- "The trainee and attending reports describe the same findings."

**Incorrect Answer**

- "Attending mentions benign mass mammographically visible, which the trainee does not include in their report" (GPT was wrong in Incosistent Findings)
- "resident does not mention the benign calcs." (GPT was wrong in Incosistent Findings)
- "attending mentions 'heterogenously dense'" (GPT was wrong in Incosistent Descriptions)
- "Discrepancy in breast density description" (GPT was wrong in Incosistent Descriptions)
- "*Complicated* cyst by itself does not lead to BI-RADS 3." (GPT was wrong in Incosistent Diagnoses)
- "BR1 appropriate if no imaging findings are identified, regardless of palpability" (GPT was wrong in Incosistent Diagnoses)

**Stylistic Differences**

- "this is stylistic"
- "The terms used in the resident and attending reports are often interchangeable. The terms are not inconsistent with one another."

**Clinical Irrelevance**

- "4A vs 4 distinction is not clinically relevant in this case."
- "There is discrepancy in distribution but not as described by GPT; relevant difference is multiple quadrants vs regional for calcifications. Low suspicion vs suspicious is not a clinically relevant discrepancy."

**Correct but Rationale Wrong**

- "Although GPT is correct, the rationale is incorrect (one term is not more specific than the other; the two are distinct categories)"
- "I don't like that GPT states that BI-RADS 3 is appropriate for "low probability of malignancy" which to me is more appropriate for BI-RADS 4A findings"

**Reader Self-Correction**

- "I missed that !"
- "that is true - I misread."

Supplementary Tables

**Table S1:** Distribution of BI-RADS categories in the 100 reports used for the reader study, across ultrasound, mammography, and overall assessments. Some reports included only an overall BI-RADS without separate ultrasound or mammography values, which accounts for discrepancies in category counts across modalities.

| BI-RADS Category | Ultrasound | Mammogram | Overall |
| --- | --- | --- | --- |
| BI-RADS 0 | 7 | 11 | 14 |
| BI-RADS 1 | 23 | 24 | 24 |
| BI-RADS 2 | 30 | 26 | 44 |
| BI-RADS 3 | 2 | 1 | 9 |
| BI-RADS 4 | – | – | 9 |

**Table S2**: Inter-reader agreement for each of the three error types, measured using Krippendorff's α with 95% confidence intervals. Agreement among human readers is reported in the "α (Readers)" column for all readers, attendings only, and residents only. The **"Δ"** column shows the change in agreement when GPT-4o replaced a human reader, along with 95% confidence intervals. None of the changes were statistically significant, as indicated by the corresponding p-values. Higher alpha values indicate stronger inter-reader agreement.

| Group | Error Type | α (Readers) | Δ | p-value |
|---|---|---|---|---|
| **All** | Inconsistent Findings | 0.767 (0.679–0.842) | -0.004 (-0.019 – 0.008) | 0.875 |
| | Inconsistent Descriptions | 0.595 (0.506–0.680) | -0.013 (-0.037 – 0.008) | 0.626 |
| | Inconsistent Diagnoses | 0.567 (0.395–0.695) | 0.002 (-0.030 – 0.028) | 0.751 |
| **Attendings** | Inconsistent Findings | 0.813 (0.717–0.895) | -0.026 (-0.071 – 0.010) | 0.748 |
| | Inconsistent Descriptions | 0.646 (0.534–0.746) | -0.084 (-0.157 – -0.018) | 0.499 |
| | Inconsistent Diagnoses | 0.648 (0.451–0.793) | -0.051 (-0.133 – 0.019) | 0.494 |
| **Residents** | Inconsistent Findings | 0.723 (0.612–0.819) | 0.008 (-0.030 – 0.040) | 0.504 |
| | Inconsistent Descriptions | 0.586 (0.471–0.690) | 0.010 (-0.040 – 0.054) | 1.000 |
| | Inconsistent Diagnoses | 0.518 (0.317–0.664) | 0.042 (-0.017 – 0.096) | 0.250 |

**Table S3**: Number of optional free-text comments left by readers regarding GPT-4o's feedback, categorized by reader group and error type. Each cell with a white background shows the total number of comments submitted. The last row (gray background) reports the number of unique report pairs (samples) that received at least one comment for each error type. For example, in Inconsistent Findings, comments were left on 34 unique samples, totaling 89 comments (since multiple readers could comment on the same sample). Note that the total in the last column of the gray row (78) reflects the number of unique samples that received at least one comment across any error type; it is not a direct sum of the per-error-type counts, as some samples received comments on multiple error types.

|  | Inconsistent Findings | Inconsistent Descriptions | Inconsistent Diagnoses | Total (Row) |
|---|---|---|---|---|
| **Attendings** | 47 | 73 | 45 | 165 |
| **Residents** | 42 | 65 | 39 | 146 |
| **All Comments** | 89 | 138 | 84 | 311 |
| **Number of Report Pairs** | 34 | 59 | 35 | 78 |

| that Received Comments | | | | |
|---|---|---|---|---|
| | | | | |

# Supplementary Figures

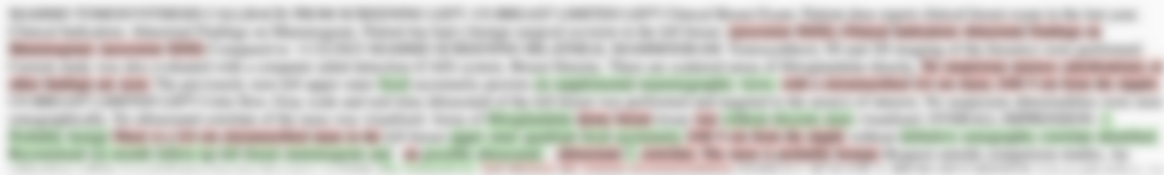

**Figure S1**: Screenshot of the labeling interface in Label Studio as seen by readers. After brief instructions, readers were shown a single clinical report with differences between the resident's draft and the attending's final version. Text added by the attending was highlighted in green, and text

removed by the attending was highlighted in red. A dropdown menu allowed access to both original report versions. Readers then answered three questions evaluating the three error types, each with an optional comment box. Once these were completed, a second set of three questions appeared, asking readers to evaluate GPT's responses, also with optional comment boxes.

| Error Type | Draft Report | Final Report | Attending's Answer | GPT's Answer | GPT's Feedback | Attending's Comment |
|---|---|---|---|---|---|---|
| Inconsistent Findings | Right 9:00, 6 cm from the nipple, anechoic mass versus cluster of cysts with indistinct margins, measuring 0.6 x 0.3 x 0.6 cm. Left 12:00, 1 cm from the nipple, cyst with indistinct margins, measuring 0.8 x 0.4 x 0.8 cm. | Benign right 9:00, 6 cm from the nipple, oval cyst measuring 0.6 x 0.3 x 0.6 cm. Left 12:00, 1 cm from the nipple, benign cluster of cysts, measuring 0.8 x 0.4 x 0.8 cm. | No | Yes | The resident's report mentions a "right 9:00, 6 cm from the nipple, anechoic mass versus cluster of cysts with indistinct margins," which is not described in the attending's report. The attending describes this as a "benign right 9:00, 6 cm from the nipple, oval cyst." Additionally, the resident's report mentions a "left 12:00, 1 cm from the nipple, cyst with indistinct margins," which is not described in the attending's report. The attending describes this as a "benign cluster of cysts." | Findings are consistent in each report, only descriptors differ |
| Inconsistent Findings | In the left breast, there is a U-shaped clip. No suspicious masses, calcifications, or other findings are seen. No significant interval changes. | There is a stable biopsy clip and a benign oval mass in the left breast. | Yes | No | The resident's report does not lack or contain a finding that the attending's report has. Both reports mention the same findings, such as the oval hypoechoic mass at the midline sternum and the stable benign mass in the left breast. | Attending mentions benign mass mammographically visible, which the trainee does not include in their report |
| Inconsistent Descriptions | The questioned asymmetry at the right medial breast, posterior depth is less conspicuous on spot compression views | The questioned asymmetry at the right medial breast, posterior depth, effaces on additional views, compatible with benign overlapping fibroglandular tissue. | No | Yes | The resident's report describes the questioned asymmetry as "less conspicuous on spot compression views," while the attending's report describes it as "effaces on additional views, compatible with benign overlapping fibroglandular tissue." The attending's description provides a more definitive conclusion, indicating an inconsistency in descriptions. | The terms used in the resident and attending reports are often interchangeable. The terms are not inconsistent with one another. |
| Inconsistent Descriptions | In the right breast at 12:00, posterior depth, there are stable punctate and layering calcifications spanning 1 cm. | There are bilateral scattered calcifications in both breasts. | Yes | No | Both reports use appropriate BI-RADS lexicon terms to describe the findings, such as "oval mass" and "cyst cluster," and there are no missing or misused descriptors. Therefore, there is no inconsistency in descriptions. | Calcifications are described differently |
| Inconsistent Diagnoses | In the left breast at 3:00 2 cm from the nipple, there is a 4 mm hypoechoic mass with sonographic features of a complicated cyst. OVERALL IMPRESSION: No imaging evidence of malignancy on the current exam(s). OVERALL BI-RADS 2: BENIGN. | N/A | No | Yes | The report assigns an overall BI-RADS score of 2 (Benign) despite the ultrasound findings suggesting BI-RADS 3 (Probably Benign) due to the presence of a complicated cyst. The BI-RADS score does not align with the findings described in the report. | Trainee report is internally consistent. A complicated cyst is a benign (BR2) finding, not BR3 as GPT implies here. |
| Inconsistent Diagnoses | No suspicious masses, calcifications, or other findings are seen. Left breast biopsy clips. Mammo BI-RADS 1: NEGATIVE | N/A | Yes | No | The report's BI-RADS score (1: Negative) is consistent with the findings described in the report, which mention no suspicious abnormalities and no evidence of malignancy. Thus, there is no inconsistency in the BI-RADS assignment. | Marker clips were described, which makes the Birads category 2 (not 1) |

**Figure S2:** Illustrative cases in which GPT-4o's feedback diverged from the attending radiologist's assessment across the three error types. Only the relevant report excerpts are shown.